\documentclass[10pt,a4paper,twoside]{article}

\usepackage{indentfirst}                        
\usepackage{graphicx}                           
\usepackage{amsgen,amsfonts,amssymb,amsbsy}     

\newenvironment{resum}{\begin{quote}\small}{\end{quote}}

\newcommand{\bfsf}[1]{\textsf{\textbf{#1}}}

\begin{document}

\thispagestyle{plain}           

\begin{center}


{\LARGE\bfsf{Initial data for black hole collisions}}

\bigskip


\textbf{Sergio Dain}$^1$


$^1$\textsl{Max-Planck-Institut f\"ur Gravitationsphysik, Albert-Einstein-Institut}

\end{center}

\medskip


\begin{resum}
  I describe the construction of initial data for the Einstein vacuum
  equations that can represent a collision of two black holes. I
  stress in the main physical ideas.
\end{resum}

\bigskip


The physical system we want to describe is a binary system of two
black holes. It is expected that lot of such systems exist in the
universe.  Moreover, these systems are expected to have the following
two properties.  First,
gravitational radiation will be emitted by them and the new detectors
will be able to measure it.  Second, these systems are completed
described by the Einstein field equations.  By this I mean that there
exist solutions of the Einstein equations that can reproduce the
gravitational wave forms emitted by them. The ultimate goal is
to find these solutions and compare the waves forms with the result of
the measurements.

One method for constructing an appropriate class of solutions of the
Einstein equation is through an initial value formulation: we give
appropriate data and then we evolved them numerically. It is not
obvious that an initial value formulation is the most appropriate way
to construct a solution for an astrophysical problem because we can
not prepare the initial conditions of an astrophysical system in the
laboratory.  It is, in principle, impossible to find the exact initial
data for a real astrophysical system. The whole program will succeed
if we are able to find some properties of the binary system that do
not depend very much on fine details of the initial data. In this
sense black holes are perhaps better candidates than ordinary stars,
since they are much simpler, they do not involve the matter equations.
For example, stationary black holes are expected to be characterized
completed only by two parameters: mass and spin; on the other hand
stationary stars can have a complicated multipolar structure.

A black hole is a region of no escape which does not extend out to
infinity. In the standard text books \cite{Hawking73} \cite{Wald84}
the definition of a black hole is made using the conformal
compactification of the space time. If the space time admit a suitable
conformal boundary at null infinity and the causal past of this
boundary is globally hyperbolic, then the black hole region is defined
as the difference between the space time and the causal past of null
infinity. The boundary of the black hole region is a null surface
called the event horizon.
\begin{figure}[htbp]
  \centering
\includegraphics[width=8cm]{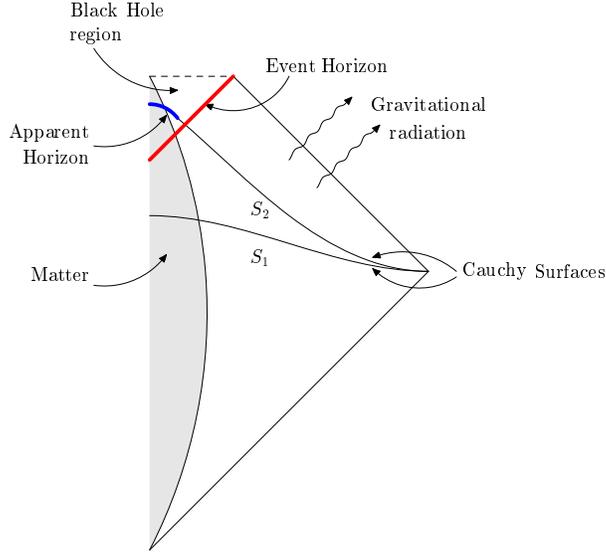}  
  \caption{Penrose diagram of a space time with a black hole.}
  \label{fig:1}
\end{figure}

I want to discuss the problem of finding initial data for black
holes. An \emph{initial data set} for the Einstein vacuum equations is given
by a triple $(S, h_{ab}, K_{ab})$ where $ S$ is a connected
3-dimensional manifold, $ h_{ab} $ a (positive definite) Riemannian
metric, and $ K_{ab}$ a symmetric tensor field on $ S$. They satisfy
the vacuum constraint equations
\begin{equation}
 \label{const1}
 D^b K_{ab} -D_a K=0,
\end{equation}
\begin{equation}
 \label{const2}
 R +  K^2- K_{ab} K^{ab}=0,
\end{equation}
on $S$, where $D_a$ is the covariant derivative with respect to
$h_{ab}$, $R$ is the trace of the corresponding Ricci tensor, and $K=
h^{ab}K_{ab}$. Examples of initial data are $S_1$ and $S_2$
in Fig. \ref{fig:1}.

We want to construct initial data set that evolve in a space time
with a black hole region. In order to do this, we must recognize the
black hole on the initial data. The event horizon is not very useful
because is a global property of the space time, we need to known the
whole space time in order to calculate it. However, there is a
remarkable consequence of the black hole theory, it is related with
the concept of the apparent horizon. An \emph{apparent horizon} can be
defined as follows. Given a slice $S$, we say that a two surface
$\Sigma$ contained in $S$ is an apparent horizon if it satisfies the
following equation
\begin{equation}
  \label{eq:1}
D_an^a-K+K_{ab}n^an^b=0,  
\end{equation}
where $n^a$ is the unit normal vector to $\Sigma$. Equation
(\ref{eq:1}) involves only quantities that are present on the initial
data. No evolution is needed. By the singularity theorems (see
\cite{Hawking73}) we know that a data with an apparent horizon will
evolve in a geodesically incomplete space time. That is, the space
time will have a singularity. If the cosmic censorship conjecture is
true, this singularity will be inside the black hole region. If there
is a black hole, then the apparent horizon must be inside the black
region. That is, the presence of an apparent horizon indicate the
presence of a black hole.

In general, a space time with a black hole will have slices with an
apparent horizon and slices without it.  In Fig. 1 the slice $S_1$
have no apparent horizon, and the slice $S_2$ contains one. Both data
describe the same space time, but looking $S_1$ it will be very
difficult to decide that the space time contain a black hole. The
presence of apparent horizon gives as a practical criteria to
distinguish a black hole data. However, it is important to recall that
in principle it is possible to have a black hole without any apparent
horizon.

A space time contains two black black hole if the event horizon has
two disconnected components. Again, is very difficult to recognize
this situation on the initial data. But, we can recognize when the
apparent horizon has two disconnected components. It seems to be
reasonable to assume that when these two components of the apparent
horizon are very separated from each other the event horizon will have
also two disconnected components.

Summarizing: a simply way of constructing a two black holes initial
data is by imposing that the data contain an apparent horizon with two
disconnected components.

The question is now how to construct initial data with apparent
horizons. In order to force the data to have an apparent horizon we
have to impose to the constraint equations (\ref{const1}),
(\ref{const2}) appropriate boundary conditions.  Black hole
data will be just a particular choice of boundary conditions.

\begin{figure}[htbp]
  \centering \includegraphics[width=8cm]{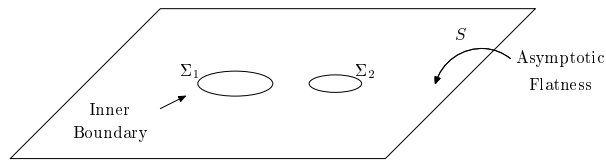}
  \caption{A black hole initial data with an apparent horizon with two
    disconnected components.}
  \label{fig:2}
\end{figure}

The boundary condition are divided naturally in two: an outer boundary
condition at infinity and an inner boundary condition. The outer
boundary condition is asymptotically flatness, it means that we are
dealing with an isolated system. It is a fall off condition for the
metric $h_{ab}$ and the extrinsic curvature $K_{ab}$. This boundary
condition is well understood. The inner boundary condition is the
condition that we have to impose on the inner boundary in order to
force it to be an apparent horizon.  The inner boundary condition is
not so well understood as the outer one.  The standard way to produce
an apparent horizon involves the choice of a non trivial topology for
the data \cite{Brill63} \cite{Misner63} \cite{Bowen80}.  These methods
introduce the apparent horizon in an indirect manner. They do not really
deal with an inner boundary. I think it would be very interesting to
know how to prescribe directly this inner boundary condition in a
consistent way.  The only article I known in this direction is a
numerical study by J. Thornburg \cite{Thornburg87}.

There exist an infinite number of data which contain apparent horizon
with two disconnected components. Which is, among them, the data that
will correspond to the real astrophysical process? If the black hole
are close to each other to answer this question is probably as
difficult as to calculate the whole space time. But if the black hole
are initially very well separated from each other then one expect
some simplifications. In particular, the first requirement we should
impose is that the data near each of the black hole should be, in some
appropriate sense, close to the data of one isolated black hole.   
We call this the \emph{far limit} of the data. The important point is
that we know  exactly the data for one black hole: the Kerr initial
data. This suggest that one way to produce an initial data with the
right far limit is by ``summing''  two Kerr initial data.

If the black holes do not have spins, then they should have a far
limit to the Schwarzschild one. The Schwarzschild initial data can be
chosen to be time symmetric, that is $K_{ab}=0$. With this assumption
the constraint equations can be reduced to a linear equation. In this
case is simpler to make sense of ``summing'' two Schwarzschild black
holes, this has been done in \cite{Brill63} \cite{Misner63}.

In the case of Kerr, the constraint equation are non linear.  It is
possible to sum part of the initial data (the ``free data''), for the
rest of the data one should solve the constraint equations. One can
prove that, under certain assumptions, these equations have solutions.
Then, one obtain an initial data which have a far limit to the Kerr
initial data. This construction  has been made in \cite{Dain00c}
\cite{Dain99b}

It important to note that in all these data the linear momentum of the
individual black holes is zero in the far limit approximation. There
exist data with arbitrary linear momentum, like \cite{Bowen80}, but
they do not have a fat limit to Kerr. They are also numerical studies
of data which have arbitrary linear momentum and also a far limit to
Kerr (see the review \cite{Cook00} and reference therein), but so far
no analytic existence proof of these type of data is available. The
main difficulty is that most of the knowledge we have about the
constraint equations is based on the assumption that the data is
maximal (i.e; $K=0$). The Kerr data in standard Boyer-Lindquist
coordinates is maximal. However, remarkably enough, is it not known
any maximal, boosted (i.e; with no trivial linear momentum), initial
data for Kerr or even for Schwarzschild. These boosted maximal slices
can in principle be used to construct a data for two black holes in
which each of them has arbitrary linear momentum in the far limit.


\end{document}